# Unprecedentedly large gap in $HgBa_2Ca_2Cu_3O_{8+\delta}$ with the highest $T_c$ at ambient pressure

Chuanhao Wen[1,5], Zhiyong Hou[1,5], Alireza Akbari[2,4,5], Kailun Chen[1,5], Wenshan Hong[3,5], Huan Yang[1✉], Ilya Eremin[2✉], Yuan Li[3], and Hai-Hu Wen[1✉]

[1]Center for Superconducting Physics and Materials, National Laboratory of Solid State Microstructures and Department of Physics, Collaborative Innovation Center for Advanced Microstructures, Nanjing University, Nanjing 210093, China

[2]Institut für Theoretische Physik III, Ruhr-Universität Bochum, D-44801 Bochum, Germany

[3]International Center for Quantum Materials, School of Physics, Peking University, Beijing 100871, China

[4]Beijing Institute of Mathematical Sciences and Applications (BIMSA), Huairou District, Beijing 101408, China

[5]These authors contributed equally: Chuanhao Wen, Zhiyong Hou, Alireza Akbari, Kailun Chen, Wenshan Hong.

✉e-mail: huanyang@nju.edu.cn; Ilya.Eremin@ruhr-uni-bochum.de; hhwen@nju.edu.cn

## Abstract

**In cuprate superconductors, the highest $T_c$ is possessed by the $HgBa_2Ca_2Cu_3O_{8+\delta}$ (Hg-1223) system at ambient pressure, but the reason remains elusive. Here we report the scanning tunneling measurements on the Hg-1223 single crystals with $T_c \approx 134$ K. The observed gaps determined from the tunneling spectra (STS) can be categorized into two groups: the smaller gap $\Delta_1$ ranges from about 45 to 70 meV, while the larger gap $\Delta_2$ from about 65 to 98 meV. The STS was measured up to 200 K and the larger gap can persist well above $T_c$, indicating a pseudogap feature which may reflect the strong pairing energy in the inner layer. Interestingly, an extremely strong particle-hole asymmetry is observed in associating with a very robust coherence-like peak at the bias of the larger gap in the hole branch of the Bogoliubov dispersion. We argue that the observed asymmetry results may be from the interplay of a flat band (van Hove singularity) in the electronic spectrum and the large gap in the underdoped (inner) layer. A theoretical approach based on a trilayer model with an interlayer coupling can give a reasonable explanation. Our results provide deep insight into understanding the mechanism of superconductivity in cuprate superconductors.**

## Introduction

Since the discovery of high-temperature superconductivity in the cuprates[1] in 1986, substantial advancements have been achieved in theoretical and experimental studies. However, the underlying mechanism of superconductivity at such high temperatures is still an enigma and continues to be an essential topic of discussion in condensed matter physics. In cuprates, *d*-wave superconductivity always coexists with the pseudogap state[2-5]. Although the origin of the pseudogap remains elusive, there are at least two potential explanations were suggested. Firstly, numerous studies have proposed that the pseudogap represents a high-temperature precursor to superconductivity[6-11]. In this scenario, preformed Cooper pairs without phase coherence have already emerged in the pseudogap state, and the scale of the pseudogap represents the pairing strength. Within the second scenario one assumes that a competing charge or spin density wave order forms below the pseudogap temperature, $T^*$, which leads to a depletion of spectral weight[4,5]. In this scenario, superconductivity emerges independently from the competing ordering at $T_c < T^*$.

Experimentally, the magnetic superexchange interaction between nearest-neighboured Cu spins may play a major role in the electron pairing[12-19]. Meanwhile, the electron-phonon coupling plays also some important roles in high-$T_c$ cuprates[20-25]. Consequently, the determining factor for the pairing strength-whether induced by phonon-mediated interaction, magnetic superexchange, or other factors, remains a challenging subject with ongoing debate. The comparison between the pairing gaps and the characteristic energies of electron-electron or electron-phonon interaction can help solve the mystery of high-$T_c$ superconductivity. To our knowledge, signatures of electron-phonon coupling were also reported in cuprate superconductors with the maximal phonon energies in different systems ranging from 50 to 80 meV, usually attributed to the optical branch of the phonon spectrum[20-24], while the superexchange energy $J$ is about 100~180 meV[15-19,26]. Thus, it is crucial to find whether the pairing gap can have an energy scale significantly greater than the largest phonon energy, manifesting that there is limited space for using the electron-phonon coupling picture to interpret the superconducting mechanism in cuprates.

Among all superconductors, the $HgBa_2Ca_2Cu_3O_{8+\delta}$ (Hg-1223) compound exhibits the highest critical temperature $T_c \approx 134$ K at ambient pressure[27] ($T_c$ reaches 164 K under pressure[28]). Like other trilayer cuprates[29], Hg-1223 has three inequivalent $CuO_2$ planes, i.e., one inner plane (IP) and two outer planes (OPs) (Fig. 1a). Previous experiments have revealed that in the trilayer Bi-2223 system, the IP may be underdoped while the OPs are overdoped[30,31]. Considering the positive correlation between the superconducting gap and the transition temperature, Hg-1223 is also expected to have a larger gap in the IP. This conclusion is strengthened by the reported gap enhancement induced by the Bogoliubov band hybridization in other trilayer cuprates[32-34]. However, due to the difficulty in preparing high-quality single crystals and the challenge of cleaving this sample, there have been no reports on Hg-1223 of studies by angle-resolved photoemission spectroscopy (ARPES) and only a few experiments using scanning tunneling microscopy/spectroscopy (STM/STS) were carried out on sintered and polycrystalline samples[35-37]. Thus, the gap size and the reason for its highest $T_c$ remain to be investigated in Hg-1223. In this article, STS measurements are conducted on optimally doped Hg-1223 samples. We observe two groups of gaps, presumably from the two bands of the IP and OPs, respectively. The largest gap can even reach a value of about 98 meV, significantly surpassing the scale of the available phonon frequencies. In addition, the intensity of the coherence peaks exhibits strong particle-hole asymmetry, and the intensity of the asymmetry has a roughly linear correlation with the larger gap size. We analyze our experimental results using trilayer mean-field superconducting model, introduced previously for Bi-2223[34], and show that the particle-hole asymmetry of the coherence-like peaks for the larger gap could be a result of interesting interplay of the gap energy in the IP with the flat band (van Hove singularity). Such effect might enhance the Cooper pairs' phase coherence leading to additional boost for the superconducting transition temperature in Hg-1223 samples compared to other trilayer cuprates. Within another scenario the second larger gap would stem from the density-wave like gap in the inner layer. Note, the charge density wave order has been reported recently in the inner layer of Hg-1223 only in the high magnetic fields[38]. Overall, our findings contribute to a deeper understanding of the superconductivity mechanism in cuprate superconductors.

## Results

## Sample characterization

A typical topographic image measured on the cleaved surface of a Hg-1223 single crystal is shown in Fig. 1b. One can see that relatively flat surfaces can be observed on both sides of a distinct step edge. The height profile, along the path indicated by the orange arrow, reveals a height drop of 15.9 Å (Fig. 1c). This value is well consistent with the $c$-axis lattice parameter[39] $c$ = 15.85 Å. Unfortunately, we failed to obtain an atomic-resolved image. Unlike in Bi-family cuprates with two Bi-O planes connected by a van der Waals bond, the Hg-1223 has only one Hg-O plane in one unit cell, which prevents achieving an atomically resolved topography. As a result, after cleavage, the exposed surface is probably reconstructed by the residual Hg and O atoms without an apparent long-range atomic order. Some clusters can also be detected on the surface, which may be formed by Hg atoms. However, the electronic structure can still be detected by the tunneling spectrum measurement although the surface is not atomically flat[40-42] and tunneling spectra with high quality can still be widely measured even with the tip traveling in a large distance on the surface in the present Hg-1223 sample.

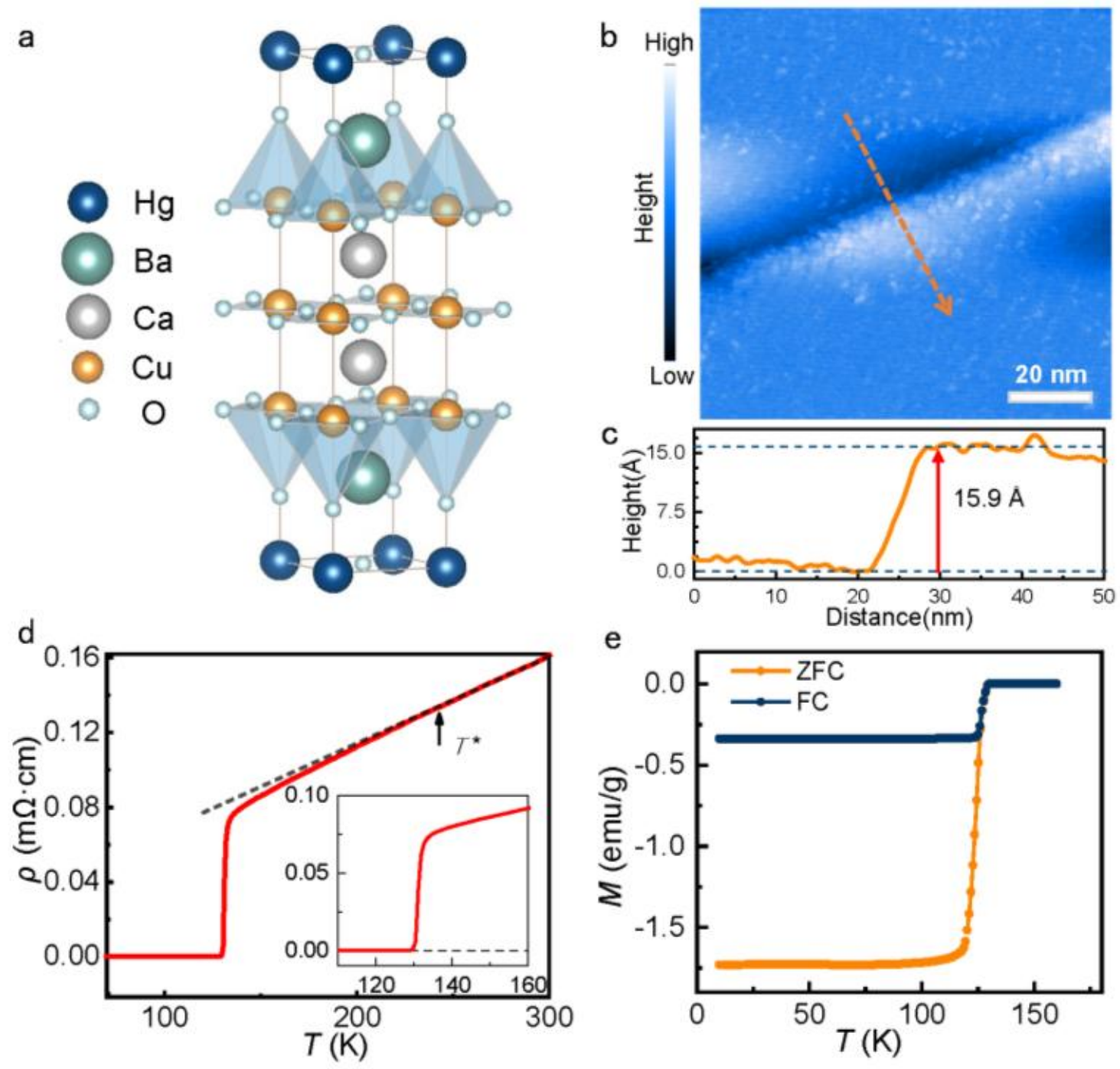

**Fig. 1| Crystal structure and sample characterization of Hg-1223. a**, Crystal structure of Hg-1223 in which two outer planes (OPs) and one inner plane (IP) are contained in a unit cell. **b**, Topography of the cleaved surface showing a single-unit-cell step. Setpoint condition: $V_{\mathrm{set}} = 1$ V, $I_{\mathrm{set}} = 20$ pA. **c**, Height profile obtained along the orange arrow shown in **b**. **d**, Temperature dependence of in-plane resistivity measured at 0 T. The dashed straight line in the main panel shows the linear extrapolation of the high-temperature resistive curve near the room temperature. The deviation temperature from the linear extrapolation and the experimental data is about 240 K, denoting the possible closing temperature of the pseudogap $T^*$. **e**, Temperature dependent magnetization measured in zero-field-cooled (ZFC) and field-cooled (FC) processes under a magnetic field of 10 Oe.

Figure 1d shows the temperature-dependent in-plane resistivity of the Hg-1223 crystal, and the onset transition temperature $T_c^{\mathrm{onset}}$ is about 134 K. Meanwhile, the resistive curve near room temperature shows a linear temperature-dependent behavior, while the curve shows a downward curvature which may be a signature of the presence of a pseudogap[43]. Therefore, we do a linear extrapolation of the resistive curve near room temperature and determine the closing temperature of a pseudogap by the deviation of the extrapolation and the experimental data. The obtained closing temperature of the pseudogap $T^*$ is about 240 K which is about 1.8 times as large as $T_c^{onset}$. We should emphasize that the determination of the pseudogap closing temperature is not precise enough, but it provides us with a rough temperature for the gap closing. Figure 1e presents the temperature-dependent magnetic susceptibility measured at 10 Oe in zero-field-cooled and field-cooled modes. Both measurements show sharp superconducting transitions, indicative of high quality of the samples.

## Unprecedentedly large gap

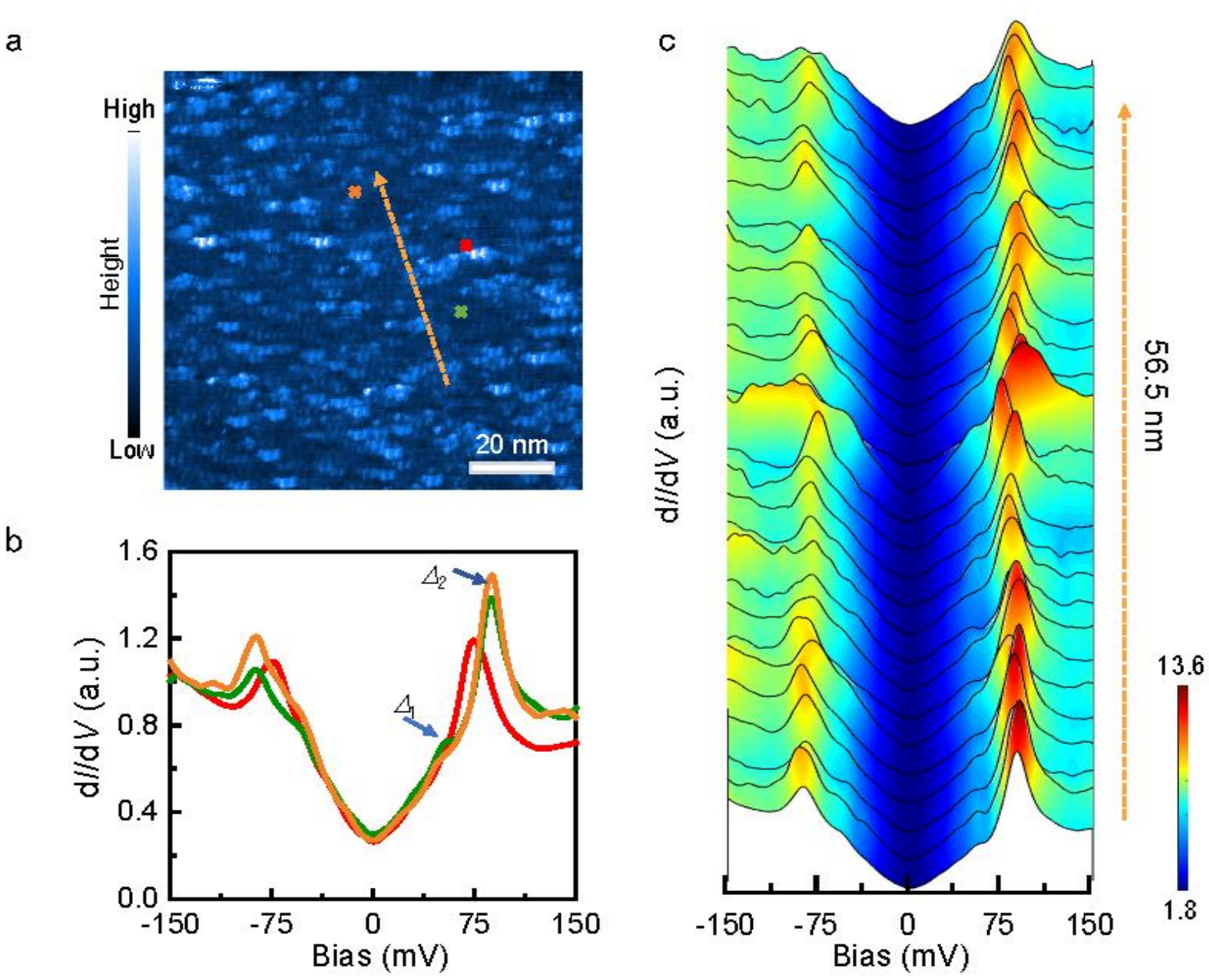

**Fig. 2| Spatially distribution of spectra and large gap. a**, STM topography of the cleaved surface. The clusters on the surface could potentially be scattered Hg/Ba atoms. **b**, Three tunneling spectra taken at the spots indicated by crosses with the corresponding colour in **a**. **c**, Spatially resolved tunneling spectra measured along the yellow dashed line in a. Each spectrum shows a two-gap feature, with the larger gap around 87meV. Setpoint condition: **a**, $V_{set}$ = 1 V, $I_{set}$ = 20 pA; **b**,**c**, $V_{set}$ = 150 mV, $I_{set}$ = 200 pA.

Then, we conduct STS measurements on the cleaved surface to study the local density of states (LDOS). In Fig. 2b, we present three representative tunneling spectra measured in the area shown in Fig. 2a. Being consistent with other cuprate superconductors[44-51], the spectra show the V-shaped feature near zero bias, which is a characteristic feature of the *d*-wave pairing gap with the function of $\Delta = \Delta_0 \cos(2\theta)$, here $\Delta_0$ is the gap maximum and $\theta$ indicates the direction in momentum space relative to the "anti-nodal" directions, which are along Cu-O plaquettes chains. The presence of non-zero differential conductance at zero bias may be due to the impurity-scattering effect in a superconductor with a nodal sign-changing gap[46,52]. After a close inspection, these d$I$/d$V$ curves show a two-gap feature, as marked by the arrows in Fig. 2b. Here we define the maximum of the smaller gap as $\Delta_1$, and the larger gap as $\Delta_2$. It is important to emphasize that the gap maximum approximates the energy of the coherence peak (kink) in a *d*-wave superconductor, especially when the scattering rate is low. Therefore, the larger gap $\Delta_2$ is characterized by the coherence-like peak at a higher bias energy on the spectrum. In contrast, the smaller gap $\Delta_1$ shows up as kinks. Strikingly, the coherence-like peak corresponding to the value of $\Delta_2$ on the positive bias exhibits a very robust intensity.

A sequence of tunneling spectra measured along the arrowed line in Fig. 2a is plotted in Fig. 2c. The two-gap feature can be seen on the spectra with different gap values. The gap inhomogeneity is an intrinsic feature in cuprate superconductors, such as $Bi_2Sr_2CaCu_2O_{8+\delta}$ and $Bi_2Sr_2Ca_2Cu_3O_{10+\delta}$ due to the defects or doping-inhomogeneity[53]. Based on Fig. 2c, the averaged gap maxima of $\bar{\Delta}_1 \approx 57$ meV and $\bar{\Delta}_2 \approx 87$ meV. These values are much larger than 20-30 meV obtained in sintered and polycrystalline samples from previous reports[36,37]. Furthermore, both the peaks and the kinks are symmetric concerning the Fermi energy. We then use the Dynes model[54] with two *d*-wave gaps to fit the tunneling spectrum (Supplementary Fig. 1). The fitting curve captures the major characteristics on the negative-bias side but deviates strongly in the positive energy region, which will be discussed later. To our knowledge, the huge gap maximum $\Delta_2 \approx 98$ meV observed in this study is the largest one ever reported, with the ratio $2\Delta_2/k_BT_c \approx 17$ which far exceeds the expected value for the BCS theory in the weak coupling limit.

## Two groups of gap values

In most places, the superconducting gaps show some inhomogeneous distribution, and an example is shown in Fig. 3a. It is evident that energies of the coherence peaks ($\Delta_2$) or kinks ($\Delta_1$) fluctuate dramatically, similar to the case in $Bi_2Sr_2CaCu_2O_{8+\delta}$ (Bi-2212) and Bi-2223[44,46,48,55]. In addition, we observe that the spectrum weights of the two gaps vary in different spectra. Some selective d$I$/d$V$ curves with two-gap features are plotted in Fig. 3b. From the bottom curve up to the top one, there is a progressive increase in the spectrum-weight proportion of $\Delta_2$, while conversely, the proportion of $\Delta_1$ exhibits a decreasing trend. Between the two extreme cases, spectra show two coherence-like peaks, which suggests a variable of contributed spectrum weight from the two gaps. The two-gap feature has also been observed on the spectra measured in another trilayer cuprate superconductor Bi-2223[36,56].

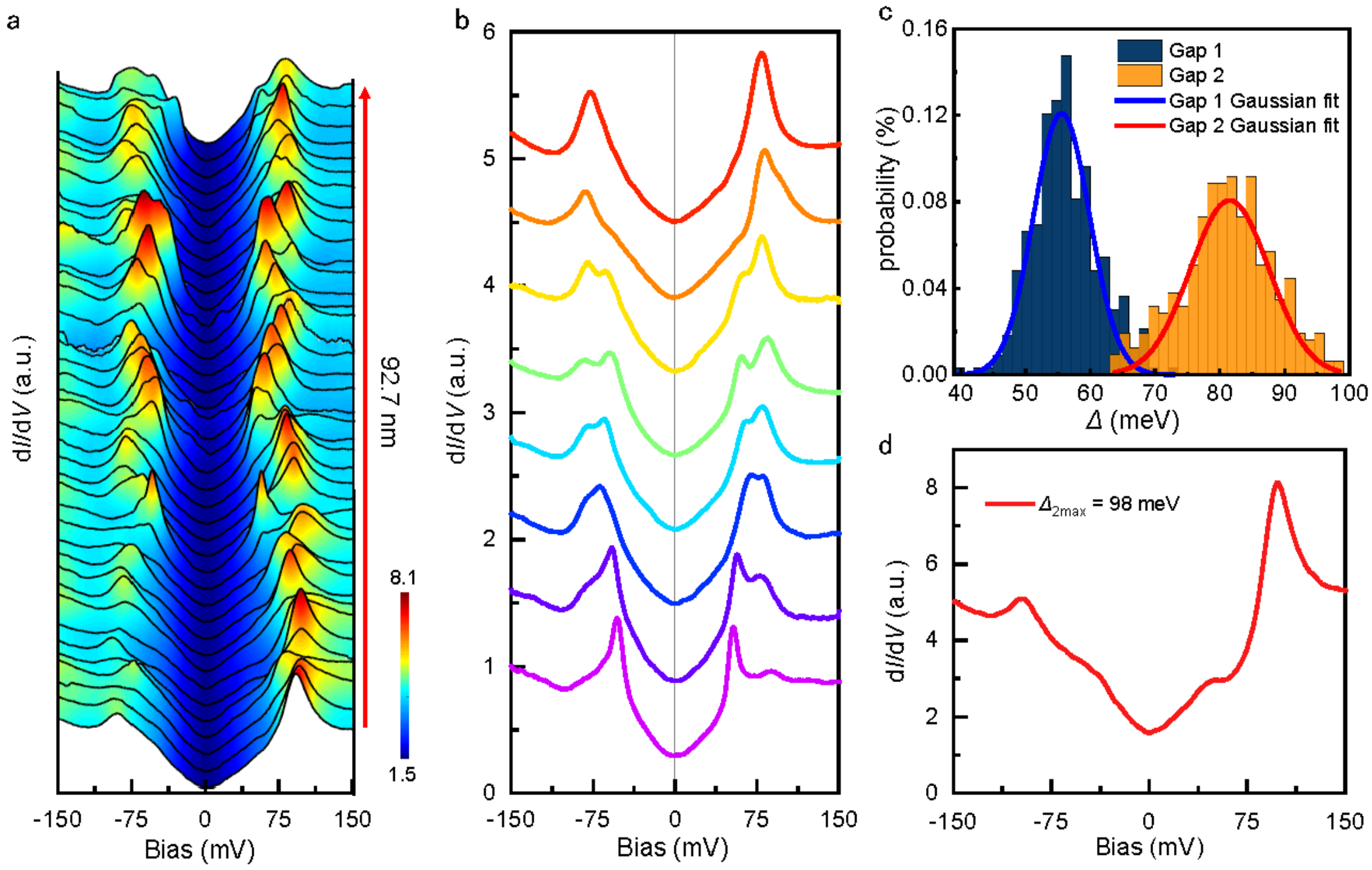

**Fig. 3| Energy distribution of the two superconducting gaps. a**, A set of tunneling spectra measured along the red line in Supplementary Fig. 2, which shows the ubiquitous inhomogeneity. **b**, Typical d$I$/d$V$ curves with different contributions of the two gaps. The spectra have been vertically offset for clarity. **c**, Histogram of the superconducting gaps. The energy distribution of the probability is fitted by two Gaussian functions (the solid lines), and the peaks of the fitting curves locate at 56 meV and 82 meV, respectively. **d**, A tunneling spectrum with the largest $\Delta_2$ of about 98 meV. The coherence-like peak at the positive bias is strong and robust. Setpoint condition: $V_{set}$ = 150 mV, $I_{set}$ = 200 pA.

Figure 3c presents a histogram showing the distribution of $\Delta_2$ and $\Delta_1$ obtained from the total of about 500 d$I$/d$V$ spectra measured in Hg-1223 samples. The distributions of the gap maxima behave as Gaussian functions for both gaps. Based on the fitting results, the peak energies are 56 and 82 meV for $\Delta_1$ and $\Delta_2$, respectively. According to a commonly perceived result[30,31], the IP and OPs are proved to be underdoped and overdoped, respectively. Therefore, it is reasonable to attribute the two gaps to these two different planes in Hg-1223.

In general, the pseudogap lacks quasiparticle-like spectral peaks and manifests solely as a suppression of electronic spectral weight. Notably, the coherence-like peaks we observed are quite sharp, particularly for the larger gap (refer to the top spectrum in Fig. 4c), which appears to differ from it. Nonetheless, numerous studies propose that the gap-feature peaks detected by STM for optimally doped and underdoped cuprates originate from the antinodal region in momentum-space and are associated with the pseudogap[57-60].

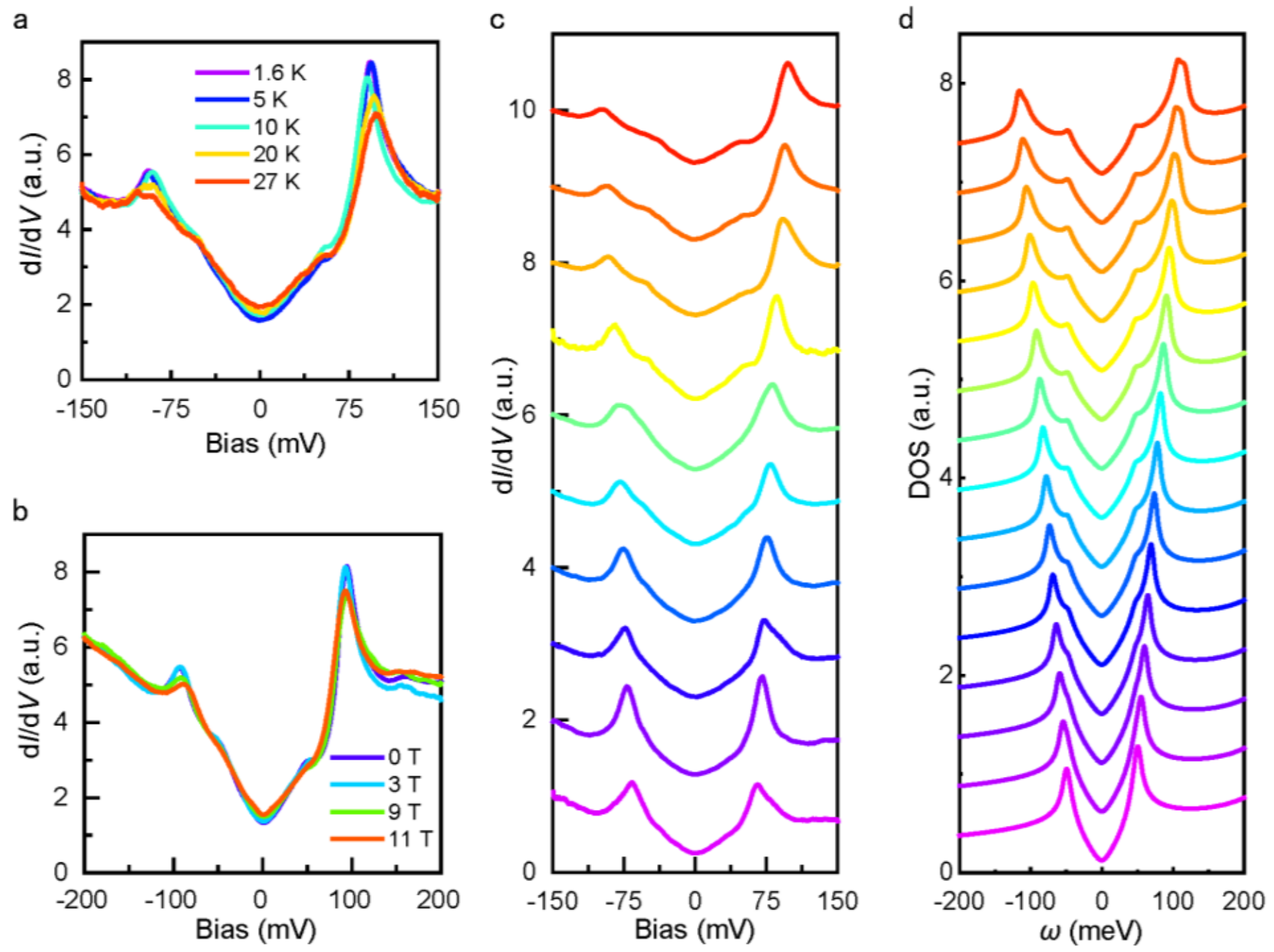

**Fig. 4| Particle-hole asymmetry in the coherence-like peaks' height. a,** Tunneling spectra measured at the same position and at different temperatures. The coherence-like peaks at $\Delta_2 \approx 93$ meV can be easily suppressed by increasing temperature. Setpoint condition: $V_{set} = 150$ mV, $I_{set} = 200$ pA. **b,** Tunneling spectra measured at different magnetic fields. The kinks and coherence-like peaks can be suppressed by the magnetic field. Setpoint condition: $V_{set} = 200$ mV, $I_{set} = 200$ pA. **c,** Tunneling spectra showing obvious particle-hole asymmetry. As the gap value increases, the positive coherence-like peak strengthens while the negative one gets suppressed. **d**, Calculated DOS spectra for the trilayer model for various values of the larger superconducting gap $\Delta_2$ (from 50 to 115 meV) by fixing $\Delta_1 = 50$ meV, $\Delta_{\mathrm{OP}}^{(0)} \approx \Delta_1$ and $\Delta_{\mathrm{IP}}^{(0)} \approx \Delta_2$. The spectra in **c** and **d** have been vertically offset for clarity.

To clarify the origin of these gaps, we examine the temperature-dependent and magnetic field evolution of the STS. The tunneling spectra obtained under high magnetic fields are shown in Fig. 4b, on these curves the spectrum weight of the larger gap emerges as the predominant component, indicate an effective suppression of both the kinks and coherence-like peaks. Furthermore, Figure 4a shows the d$I$/d$V$ curves of the same position measured at different temperatures. With the increase of temperature, the height of the coherence peaks corresponding to $\Delta_2$ is steadily suppressed, and the zero-bias differential conductance rises concurrently. Due to thermal broadening effects, the gap value seems to get larger at higher temperature, which can also be found in previous studies[57,59] of Bi-2212.

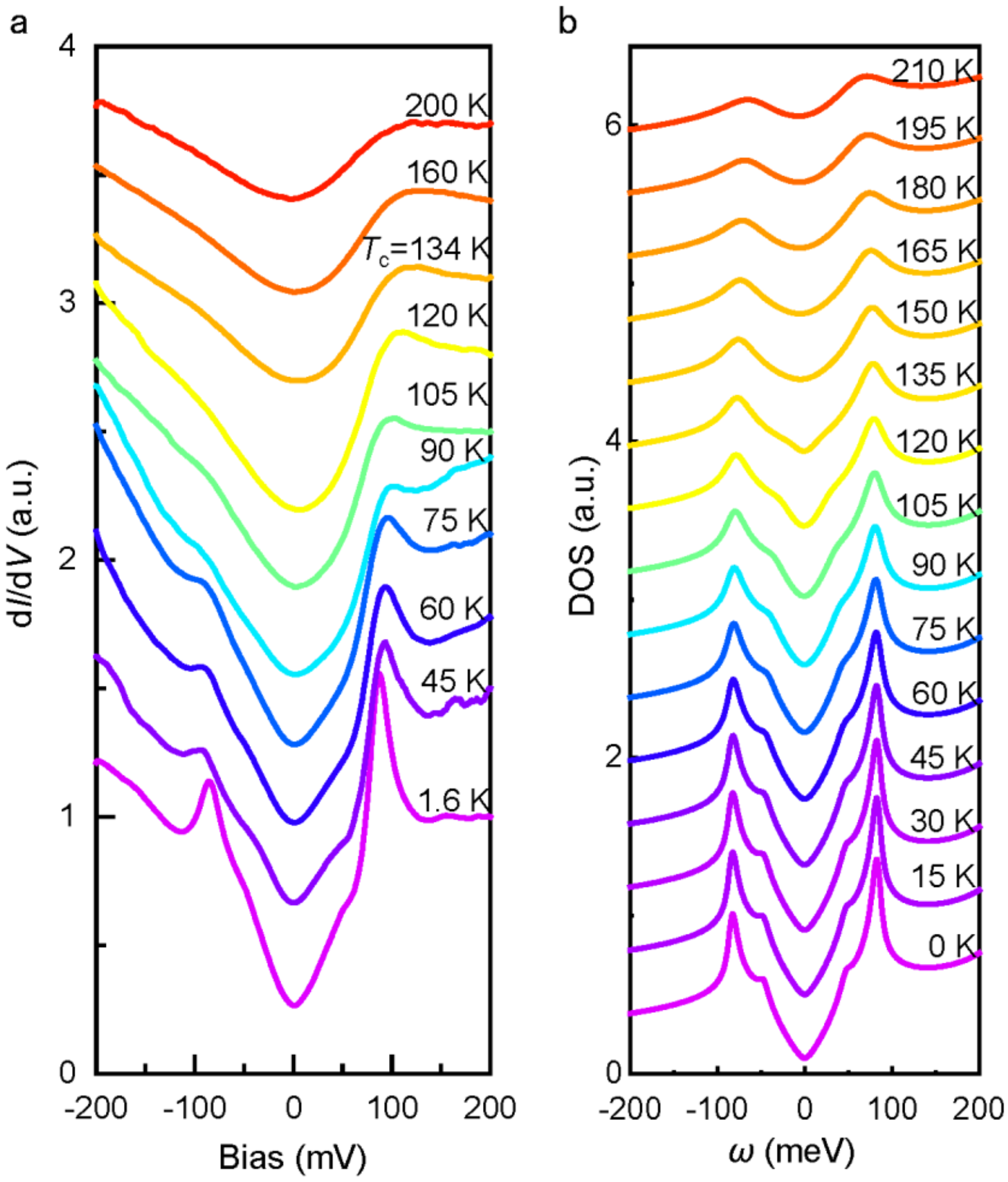

**Fig. 5 | Temperature dependence of the gap-feature in a wide range across $T_c$. a,** Averaged tunneling spectra from locations with gap value $\Delta_2$ around 88 meV at various temperatures. The kink-feature disappears during 95-105 K, while the large gap can persist up to 200 K. The spectra have been normalized at 200 mV and vertically offset for clarity. Setpoint condition: $V_{set}$ = 200 mV, $I_{set}$ = 200 pA. (b) Calculated DOS spectra for the trilayer model as a function of temperature. The temperatures are set from T~0 (bottom) to T = 210K (top) by the step of 15K. Here we assume the smaller gap, $\Delta_1$ closes at $T_c$ and the larger gap, $\Delta_2$ closes at 2$T_c$ and both follow standard BCS-like temperature dependence. In addition, we further assume some thermal broadening of the quasiparticles, which increases with temperature.

Considering that the pseudogap can also be influenced by magnetic field and increasing temperature, it is necessary to determine whether the gaps close at $T_c$ (according to the definition of the pseudogap). Due to the significant thermal drift, we were unable to track the same position during a wide range of temperature change. To avoid the influence of gap-inhomogeneity, we show the averaged tunneling spectra from locations with the similar gap value at higher temperatures in Fig. 5a. As one can see, the kink at $\Delta_1$ exhibits a stronger temperature dependence, merging into the background quickly as the temperature rises and eventually disappears at 105 K. Based on this, we can safely conclude that the small gap $\Delta_1$ is a superconducting gap. In contrast, although the coherence-like peaks corresponding to $\Delta_2$ also diminish in intensity with increasing temperature and vanish at $T_c$, the gap-feature persists up to 200 K which is consistent with gap closing temperature $T^* \approx 240$ K roughly determined from the resistive measurement. The spectra obtained above $T_c$ exhibit similar characteristics to those from the normal state of underdoped Bi-2212[58]. It indicates that the larger gap $\Delta_2$ from the underdoped IP may be a pseudogap, despite its robust coherence-like peaks at lower temperatures.

## Strong particle-hole asymmetry

As previously mentioned, the Dynes model fitting curve shows a deviation in the positive energy region, particularly around the coherence peak, see Supplementary Fig. 1. This deviation can be attributed to the fact that the larger gap has a stronger coherence-like peak at the positive bias. From our experiments, the coherence peak intensity at the positive energy side ($P_+$) is clearly higher than that at negative side ($P_-$), but the energy values of the coherence peaks are symmetric about the Fermi level. Asymmetric tunneling spectra have been previously observed in cuprate superconductors[61,62], but most of the asymmetry behaves as a higher coherence-peak intensity at negative energies than the positive energies[46,49,63]. Here in Hg-1223, the opposite situation occurs. To illustrate this asymmetry, we present one spectrum with the largest gap in Fig. 3d, and $\Delta_2$ is as large as 98 meV. A remarkably strong and robust coherence peak is evident in the hole branch of the Bogoliubov dispersion. In contrast, the $P_-$ is notably lower in comparison. Moreover, the spectrum measured on a broader energy window shows that the background DOS at negative bias is even larger than that of positive energy, which consists with previous results which attributed this asymmetry to the correlation effect, while the observed particle-hole asymmetry at the larger gap $\Delta_2$ is strongly tied to the coherence peak, and should not be a result of doping-induced correlation changes[63], but rather an intrinsic electronic property of the studied system. Next, we focus on the impact of gap size on particle-hole asymmetry. Figure 4b shows a collection of some selected and normalized spectra, each corresponding to a different $\Delta_2$ value ranging from 65 to 98 meV. A clear trend emerges as the gap size $\Delta_2$ increases: $P_+$ gets stronger, while $P_-$ experiences a gradual suppression. To provide a quantitative description of this observed trend, we define a physical parameter, $\Delta P/P = (P_+ - P_-)/(P_+ + P_-)$, to represent the normalized difference between the intensities of coherence peaks at positive and negative biases. Supplementary Fig 4a summarizes the asymmetry parameter $\Delta P/P$ of the larger gap as a function of $\Delta_2$, and the data are shown as coloured hexagons. More data have been extracted from spectra with different gap sizes, and they are presented as blue squares in Supplementary Fig. 4a. Interestingly, $\Delta P/P$ demonstrates a positive and almost linear correlation with $\Delta_2$. As $\Delta_2$ increases, the particle-hole asymmetry of coherence-like peaks becomes more pronounced. When the same statistical analysis is applied to $\Delta_1$ (Supplementary Fig. 4b), it is found that most values of $\Delta P/P$ are negative, consistent with the background energy-dependent DOS (Supplementary Fig. 3). Moreover, $\Delta P/P$ for $\Delta_1$ shows a weaker and random distribution and seems to be independent of $\Delta_1$.

## Model calculation

To model the three-$CuO_2$-layer system we employ a tight-binding model for coupled $CuO_2$ planes, which has been previously used for Bi-2223 systems[34] and its details is presented in the Methods section. We note here that due to the absence of high-quality ARPES data at present in this compound one could use various set of the tight-binding parametrizations to find the reasonable fit to the experimental data. In particular, here we assume that the OP and IP bands are coupled primarily through the single-particle hopping between the OPs and IP $t_{OO}(\boldsymbol{k}) = t_{OO}$ and $t_{IO}(\boldsymbol{k}) = t_{IO}\left(\cos k_x - \cos k_y\right)^2$, which are similar to the values found for Bi-2223 compound[34]. The hybridizations between the planes are given by $t_{OO}(\boldsymbol{k}) = t_{OO}$ and $t_{IO}(\boldsymbol{k}) = t_{IO}\left(\cos k_x - \cos k_y\right)^2$ and we set $t_{OO}$ = 17.0 meV, and $t_{IO}$ =8.5 meV. We further assume that the intralayer electronic dispersions within each layer has the form $\varepsilon_i(\boldsymbol{k}) = t_{i0} + 2t_{i1}\left(\cos k_x + \cos k_y\right) + 4t_{i2}\cos k_x \cos k_y + 2t_{i3}\left(\cos 2k_x + \cos 2k_y\right)$, where $t_{i0}, t_{i1}, t_{i2}, t_{i3}$ are the chemical potential, in-plane nearest neighbor, second-nearest-neighbor and third-nearest-neighbor in-plane hoppings, respectively. We choose the hopping integrals $(t_{O0}, t_{O1}, t_{O2}, t_{O3})$ = (136, -221, 85, -8.5) meV for the outer (overdoped) planes and $(t_{I0}, t_{I1}, t_{I2}, t_{I3})$ = (132.6, -119, 34, -5.95) meV for the inner (underdoped) plane. Most importantly, the parametrization is chosen such that the flat band for the underdoped layer at $\mu_i^{VHS} = -t_{i2} + t_{i3} + t_{i0}$ is located at small negative bias, i.e., van Hove singularities (VHS) are located within the energy window of $[-\Delta_i, \Delta_i]$ with respect to the Fermi level in the normal state as shown in Supplementary Fig. 5a. The superconducting gaps for the IP and OPs have a *d*-wave form $\Delta_{i\boldsymbol{k}} = \frac{\Delta_i^{(0)}}{2}(\cos k_x - \cos k_y)$. From previous data on Bi-2223 it is known that the interlayer Cooper-pairing tunneling gap $\Delta_{\boldsymbol{k}}^{\text{inter}}$ is also of *d*-wave structure, dictated by symmetry but its magnitude is at most 10% of the intralayer gaps[34] and we checked that its inclusion does not change the results in a significant manner. The resulting Bogoliubov energy dispersion are shown in Supplementary Fig. 5. The Green's function matrix for a superconducting state is defined as $G(\boldsymbol{k}, i\omega) = \left[i\omega - \hat{H}_{\text{SC}}\right]^{-1}$ and the generalized density of states (DOS) is calculated in the continuum limit $\rho(\omega) = -\frac{1}{\pi} Im[\sum_{\boldsymbol{k}} Tr'[G(\boldsymbol{k}, i\omega)]]_{i\omega \to \omega + i0^+}$. To understand better the interplay between the superconducting gap magnitudes of OPs and IP and van Hove singularities we plot in Supplementary Fig. 6 the evolution of the DOS for varied magnitudes of the $\Delta_{\text{OP}}^{(0)}$ and the $\Delta_{\text{IP}}^{(0)}$ assuming for a fixed gap $\Delta_1$ = 55 meV (inferred from the central value of experimental data, Fig. 3c), and $\Delta_2$ is varied from 50 to 120 meV. As we show in the Supplementary Note 1 for the single-band case the density of states of the Bogoliubov quasiparticles will show two features: the particle-hole symmetric coherence peaks at $\omega \approx \pm \Delta_i$ and the particle-hole *asymmetric coherence-like* peaks at $\omega \approx \pm \sqrt{\left(\mu_i^{VHS}\right)^2 + \Delta_i^2}$ and the ratio of the asymmetry is determined whether $\mu_i^{VHS}$ in the normal state is initially located at the positive or negative bias. For the negative bias, i.e. $\mu_i^{VHS} < 0$, the coherence-like peak is larger at positive bias and visa versa. In the situation when $\left|\mu_i^{VHS}\right| \ll \Delta_i$ the two features merge into a single particle-hole asymmetric coherence-like peaks and the asymmetry is increasing the larger $\Delta_i$ becomes. Note further that the single-band results would not explain the evolution of the experimental data and require an inclusion of the entire three-bands structure. For a chosen tight-binding parametrization in the three-bands case, the position of the van Hove singularity in the inner (underdoped) layer is located at the small negative energy in the normal state at around $\mu_2^{\text{VHS}} \approx$–16 meV<< $\Delta_2$ and the resulting density of state shows a strong particle-hole asymmetry at $\omega \approx \pm\sqrt{(\mu_2^{VHS})^2 + \Delta_2^2}$, which is slightly larger than $\omega \approx \Delta_2$. The evolution of the density of states with various values of $\Delta_2$ is further shown in Supplementary Fig. 6. Most importantly, we find that the density of states shows three well-distinguishable features: two coherence peaks at $\omega_1 \approx \pm \Delta_1$ and $\omega_2 \approx \pm \Delta_2$ as well as particle-hole asymmetric coherence-like peaks at $\omega_3 \approx \pm \sqrt{\left(\mu_2^{\mathrm{VHS}}\right)^2 + \Delta_2^2}$. Note that the exact values depend on the shape of the Fermi surface for a fixed value of $\Delta_i$ and the strength of the hybridization between different layers. The larger magnitude of $\Delta_2$ is, the stronger is the particle-hole asymmetry of the peaks at $\omega_3$ in qualitative agreement with the experimental data. To make a direct comparison to the experimental result we present the simulated data in the similar fashion as it is done in the experiment, see Fig. 4d in the main text. We further notice here that for detailed quantitative comparison further information on the electronic structure from the experiment would be needed, which is not available at present. At the same time, we believe that the presence of the flat band near the Fermi level in the trilayer Hg-1223 could be of significance to understand the origin of the highest $T_c$ in this compound and deserves further investigation. We should emphasize that this theoretical model may be a possible reason for the particle-hole asymmetry in coherence-peak heights, and the experimental data require more theoretical supports.

## Discussion

One of the most important issues is to find out what determines the critical temperature of a cuprate superconductor. Here, the most puzzling point is what characteristic enables Hg-1223 to host the highest critical temperature. Within cuprates, it is widely recognized that $T_c$ is influenced by the number of $CuO_2$ planes ($n$) per unit cell, and it peaks at $n = 3$ and begins to decline with a further increase of the number[64] $n$. In Bi-2223, enhanced gaps induced by Bogoliubov band hybridization between IP and OPs are observed[21-23], which may be the cause of the relatively high $T_c$. At the same time, the interlayer gaps were found relatively moderate i.e. of the order of 10% of the intralayer gaps[32-34]. Given that Hg-1223 is also a trilayer cuprate, this phenomenon is expected to occur as well. Moreover, the measured superexchange energy $J$ of Hg-1212 is larger than that of Bi-2223[19]. Consequently, it is logical to anticipate that Hg-1223, with higher $T_c$, will possess a larger $J$ and, furthermore, will have a larger gap size as evidenced by our observations. Concerning the origin of the larger gap $\Delta_2$, we note that it exists well above $T_c$. This is a characteristic signature of a pseudogap. In our analysis we followed the proposal that it is of superconducting nature, originates from the IP layer and represents the so-called pairing energy scale without the phase coherence. We were able to describe its evolution with temperature consistently with experimental data, assuming that the larger gap at $\Delta_2$ closes at the inner layer at temperature of about $2T_c$. We cannot completely exclude that $\Delta_2$ is connected to the density wave order yet we believe it is unlikely as we see no signature of density wave transition above $T_c$ from the resistivity data. Given its superconducting origin, the observed unprecedentedly large gap may explain the highest $T_c$ in the Hg-1223 system. Furthermore, the energy scale of $\Delta_2$ exceeds the largest value of the phonon frequencies ever reported in all cuprate superconductor[20-24]. This may exclude the possibility of interpreting the superconductivity based on a phonon mediated picture.

Another surprising aspect we observed in our study is the strong particle-hole asymmetry of coherence peaks associated with the larger gap $\Delta_2$ from the IP layer, although particle-hole asymmetry of coherence peaks was also reported in Bi-2223 and was attributed to the pair-breaking scattering between flat antinodal Bogoliubov bands[65]. As follows from our analysis the particle-hole asymmetry of the coherence-like peaks in Hg-1223 may stem from the interplay of the VHS singularity from the normal state electronic spectrum with the gap energy in the IP layer, which points towards another possible mechanism of $T_c$ enhancement in these trilayer systems. This is certainly beyond BCS paradigm and requires further analysis. Previous ARPES studies have revealed that the height of superconducting coherence peaks is proportional to the superfluid density[66,67]. This complicated electronic structure enables Hg-1223 to possess considerable pairing strength and strong phase stiffness at the same time, which may account for the highest $T_c$. Given the new insight, a promising way to further enhance $T_c$ could be to band-engineer the location of VHS relative to the Fermi level, so that its interplay with the superconducting gap can help boost the phase coherence of Cooper pairing.

## Acknowledgements

We acknowledge helpful discussions with Tao Xiang at IOP, CAS and Qianghua Wang at NJU. This work was supported by the National Natural Science Foundation of China (Grant Nos. NSFC-DFG/12061131001, 11927809, 12061131004, 11888101) and National Key R&D Program of China (Grant No. 2022YFA1403201, 2021YFA1401900). The work of I. E. and A. A. is supported by the

German Research Foundation within the bilateral NSFC-DFG Project No. ER 463/14-1. W. H. acknowledges support from the Postdoctoral Innovative Talent program (BX2021018) and the China Postdoctoral Science Foundation (2021M700250).

## Author contributions

Single crystals were synthesized by W.H. and Y.L. STM/STS measurements and analysis were performed by C.W., Z.H., K.C., H.Y. and H.H.W. The theoretical calculation was done by A.A. and I.E. H.H.W., H.Y., C.W., and I.E. wrote the paper. H.H.W. coordinated the whole work. All authors have discussed the results and the interpretations.

## Competing interests

The authors declare no competing interests.

## Methods

**Sample preparation and characterization.** Optimally doped $HgBa_2Ca_2Cu_3O_{8+\delta}$ single crystals were grown with a self-flux method[68]. Temperature-dependent resistivity measurements were carried out with a physical property measurement system (PPMS-9T, Quantum Design). The DC magnetization measurements were performed with a SQUID-VSM-7T (Quantum Design).

**STM/STS measurements.** The STM/STS measurements were carried out in a scanning tunneling microscope (USM-1300, Unisoku Co., Ltd.) with ultra-high vacuum, low temperature, and high magnetic field. The samples were cleaved in an ultra-high vacuum with a base pressure of about $1 \times 10^{-10}$ torr at liquid nitrogen temperatures. The electrochemically etched tungsten tips were used for the STM/STS measurements. A typical lock-in technique was used for the tunneling spectrum measurements with an AC modulation of 3mV and 931.773 Hz. The offset bias voltages in STS measurements have been carefully calibrated. The STM/STS measurements are carried at 1.6 K except for the variable temperature measurements.